\documentclass[ reprint, amsmath,amssymb,aps]{revtex4-1}

\usepackage{graphicx}
\usepackage{dcolumn}
\usepackage{bm}
\usepackage{xcolor}
\newcommand{\bea}{\begin{eqnarray}\displaystyle}
\newcommand{\eea}{\end{eqnarray}}
\newcommand{\nn}{\nonumber}

\newcommand{\figref}[1]{Fig.~\protect\ref{#1}}

{\setlength{\fboxsep}{15pt}
\setlength{\mylength}{\linewidth}%
\addtolength{\mylength}{-2\fboxsep}%
\addtolength{\mylength}{-2\fboxrule}%
\Sbox
\minipage{\mylength}%
\setlength{\abovedisplayskip}{0pt}%
\setlength{\belowdisplayskip}{0pt}%
\equation}%
{\endequation\endminipage\endSbox
\[\fbox{\TheSbox}\]}
\begin{document}

\title{Compactified Webs and Domain Wall Partition Functions}

\author{ Khurram Shabbir}
 \vskip 2cm

\affiliation{Department of Mathematics, Government College University, Lahore, Pakistan.}%

\date{\today}

\begin{abstract} In this paper we use the the topological vertex formalism to calculate a generalization of the ``domain wall" partition function of M-strings. This generalization allows calculation of partition function of certain compactified webs using a simple gluing algorithm similar to M-strings case.\end{abstract}

\maketitle  

\tableofcontents
\section{Introduction}
In this paper we introduce higher rank M-string domain wall partition function using the topological vertex formalism \cite{AKMV}. Just like the domain wall partition function calculated in \cite{Haghighat:2013gba} these higher rank generalizations allow the calculation of partition functions of certain compactified webs as well as the M-string in the orbifold background \cite{Haghighat:2013tka, Hohenegger:2013ala} using simple gluing rules.

The M-string partition functions discussed in \cite{Haghighat:2013gba} were generalized to the case when the space transverse to the M5-branes was an $A_{N-1}$ orbifold \cite{Haghighat:2013tka, Hohenegger:2013ala, Hohenegger:2016eqy}. It was shown that these brane configurations were dual to a configuration of type IIB $(p,q)$ 5-branes in which there were $M$ D5-branes and $N$ NS-5branes where $M$ was the number of M5-branes. The partition function of these $(N,M)$ D5/NS5-branes were studied in detail in 
\cite{Hohenegger:2013ala,Hohenegger:2015cba,Hohenegger:2015btj,Hohenegger:2016eqy}. It was shown in \cite{Hohenegger:2013ala} that these partition functions can be calculated using the refined topological vertex formalism as well as using equivariant integration over the product of instanton moduli spaces. The instanton contribution to the gauge theory partition function is engineered in topological string by the contribution of holomorphic curves in certain homology classes of the Calabi-Yau threefold, depending on the instanton number. The topological vertex \cite{AKMV} and refined topological vertex \cite{IKV} formalism allow exact computation of the gauge theory partition function if the corresponding Calabi-Yau threefold is toric. In the topological vertex formalism the topological string partition function is given by sums over functions of Young diagrams and a direct connection with Nekrasov's instanton calculus \cite{Nekrasov} arises since these Young diagram label the fixed points on the instanton moduli spaces. The topological string partition function can then by understood as representing equivariant integral over the instanton moduli space.

We will show that there are certain  interesting compactified webs which are direct generalization of M-string webs and are dual resolved $\mathbb{Z}_{N}$ orbifold of the M-string Calabi-Yau threefold.  We calculate these higher rank domain wall partition functions, denoted by $G^{(N)}_{\lambda\mu}$ using the topological vertex. 


The paper is organized as follows. In section 2, we discuss the compactified brane configurations which generalize the M-strings case and show that these can be obtained from generalized domain walls which can be considered as orbifold of the domain walls of the M-strings case.  In section 3, we calculate the partition function of these higher rank domain walls using the topological vertex. In section 4, we summarize our conclusions and discuss some open questions.

\section{Compactified Webs and Domain Walls}

The duality between $(p,q)$ 5-brane webs and toric Calabi-Yau geometries \cite{LV} has led to geometric engineering of various different gauge theories and little string theories. Much has been understood about different aspects of the gauge theories from these two different yet dual ways of realizing the gauge theories in string theory. 

In this section we discuss the compactification of web diagrams which will lead to webs generalizing the 5-branes dual to the M-strings brane configuration. Recall that the ${\cal N}=1$ 5D $SU(N)$ theory with Chern-Simons coeffiient $k$ can be engineered using M-theory compactification on a Calabi-Yau threefolds, which we will denote by $X_{N,k}$ \cite{Intriligator:1997pq}. $X_{N,k}$ are toric Calabi-Yau threefolds given by a resolved $\mathbb{C}^{2}/\mathbb{Z}_{N}$ fibered over $\mathbb{P}^1$,
\bea
X_{N,k}\sim \widetilde{\mathbb{C}^{2}/\mathbb{Z}_{N}}\times_{f} \mathbb{P}^1\,.
\eea
The integer $k$ determines the details of the fibration since there are $N$ distinct fibrations. The  compact divisors in $X_{N,k}$ are $\mathbb{P}^{1}$ bundles over $\mathbb{P}^1$ known as the Hirzebruch surfaces $\mathbb{F}_{m}$. The area of the base, which we will denote by $t_b$, gives the gauge coupling of the theory and area of the various $\mathbb{P}^1$'s in the fiber (coming from the resolution of the $\mathbb{Z}_{N}$ orbifold) are related to the Coulomn branch parameters (vev of the scalars) in the theory. If $\langle \Phi\rangle=\mbox{diag}(a_{1},a_{2},\cdots, a_{N})$ is the vev of the scalar with Coulomb branch parameters $a_{i}$ (with $\sum_{i}a_{i}=0$) then the area of $(N-1)$ curves in the fiber are given by $t_{f_{i}}=a_{i}-a_{i+1}$, $i=1,2,\cdots, N-1$.

Recall that there is a duality between the toric Calabi-Yau threefolds and certain $(p,q)$ 5-brane configurations in type IIB \cite{Aharony:1997bh,LV}. The 5D theory one obtains via M-theory compactification on a toric Calabi-Yau threefold can also be obtained on the worldvolume of a set of intersecting $(p,q)$ 5-branes in type IIB. Let us denote by $x^{a},a=0,1,2,\cdots,9$ the spacetime coordinates in type IIB. We consider a set of $(p,q)$ 5-branes whcih have $x^{0},x^{1},x^{2},x^{3},x^{4}$ as common directions in their worldvolume and are oriented in various directions (are strainght lines) in the $x^{5}$-$x^{6}$ plane. The requirement of supersymmetry forces the $(p,q)$ brane to be oriented in the $(p,q)$ direction in the $x^{5}$-$x^{6}$ plane. If all the branes are of the same charge then the worldvolume theory has 16 supercharges (the brane configuration  breaks half of the 32 supercharges). However, if there are branes of different charges then there is a furthere breaking leaving 8 preserved supercharges giving an ${\cal N}=1$ theory on the worldvolume. $(p,q)$ strings ending on this web of 5-branes are then identified with the cycles of the dual Calabi-Yau threefold. Given a toric Calabi-Yau threefold the toric data is encoded in the Newton polygon and the graph dual of the Newton polygon in the web digram of the 5-branes \cite{Aharony:1997bh,LV}. The web diagram can be understood directly from the Calabi-Yau perspective as the degeneration loci of the $T^3$ fibration over the base in the sense of SYZ fibration \cite{Strominger:1996it}. We will see later that certain brane configurations in the $x^{5}$-$x^6$ plane are symmetric in such a way that they allow the $x^5$-$x^6$ plane to be rolled into a cylinder or a torus. 

\begin{figure}[h]
  \centering
  \includegraphics[width=5cm]{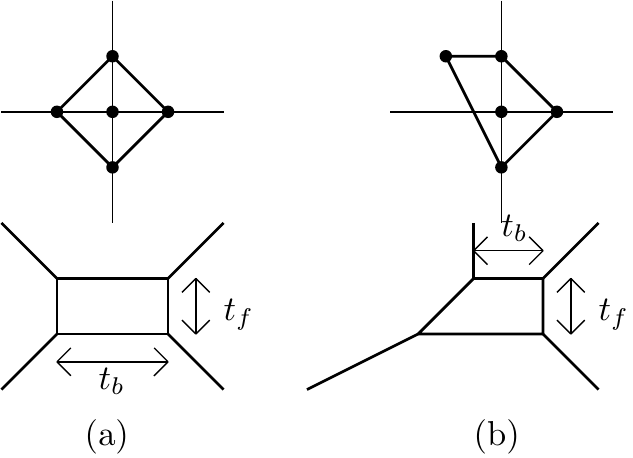}\\
  \caption{(a) Newton polygon and web of canonical bundle on $\mathbb{P}^{1}\times \mathbb{P}^1$ which gives rise to $SU(2)$ gauge theory with zero theta angle. (b) Newton polygon and web of canonical bundle on $\mathbb{F}_{1}$ which gives rise to $SU(2)$ gauge theory with theta angle $\pi$.}\label{su2webs}
\end{figure}

For the case of $N=2$ the integer $k\in \{0,1\}$ and the corresponding Calabi-Yau threefolds are the total space of the canonical bundle over the Hirzebruch surfaces $\mathbb{F}_{k}$ which are $\mathbb{P}^{1}$ bundles over $\mathbb{P}^1$ with $\mathbb{F}_{0}=\mathbb{P}^{1}\times \mathbb{P}^1$. The web diagrams and the Newton polygons corresponding to these are shown in \figref{su2webs}.

\begin{figure}[h]
  \centering
  \includegraphics[width=5cm]{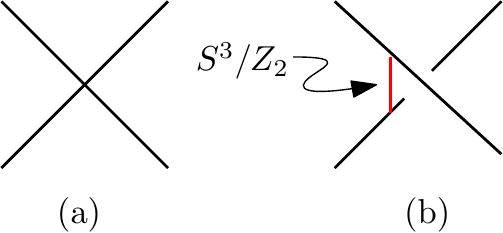}\\
  \caption{(a) Intersecting 5-branes representing a singularity in the corresponding Calabi-Yau threefold. (b) Separating the 5-branes in the transverse space corresponds to geometric transition.}\label{geometrictransition}
\end{figure}

Note here that for the case of $N=2$ the gauge group is $SU(2)$ and there is no Chern-Simons term. What distinguishes the field theory coming from these Calabi-Yau threefolds with different $k$ is the 
$\mathbb{Z}_{2}$ valued theta angle \cite{Douglas:1996xp}. Notice that in both cases the the compact divisor can be shrunk to zero size (the compact 4-cycle is the polygon in the web diagram and BPS states coming from M5-brane wrapping the 4-cycle are dual in the web to BPS states coming from D3-branes suspended between the 5-branes "filling" the polygon in the web), however, in the case of $X_{2,0}$ (the canonical bundle on $\mathbb{P}^{1}\times \mathbb{P}^{1}$) the singularity generated by the shrunken compact divisor can be deformed to get a three cycle. This transition from compact 4-cycle to singularity and resolution giving a three cycle is known as geometric transition. The three cycle in this case is $\mathbb{S}^3/\mathbb{Z}_{2}$. This is the case of $\mathbb{Z}_{2}$ action on the conifold. Recall that the deformed conifold is given by 
\bea
z_{1}z_{2}-z_{3}z_{4}=\mu\,,
\eea
with base $S^3$ given by the real locus $z_{1}=\overline{z_{2}},z_{3}=-\overline{z_{4}}$,
\bea
|z_{1}|^2+|z_{3}|^2=\mu\,,
\eea 
with $\mu$ determining the size of the $S^3$. The $\mathbb{Z}_{N}$ action (we consider the arbitrary $N$ case, the Calabi-Yau threefold $X_{2,0}$ corresponds to $N=2$) is given by,
\bea
(z_{1},z_{2},z_{3},z_{4})\mapsto (e^{\frac{2\pi i}{N}}z_{1},e^{-\frac{2\pi i}{N}}z_{2},e^{\frac{2\pi i}{N}}z_{3},e^{-\frac{2\pi i}{N}}z_{4})\,.
\eea 
In terms of the brane webs we can see that when the compact cycle represented by the rectangle in \figref{su2webs}(a) shrinks to zero size we have a $(1,1)$ 5-brane intersecting a $(-1,1)$ 5-brane. The intersection can be smoothed out by separating the branes in the transverse space as shown in \figref{geometrictransition}.
\vskip 0.5cm

\begin{figure}[h]
  \centering
  \includegraphics[width=2in]{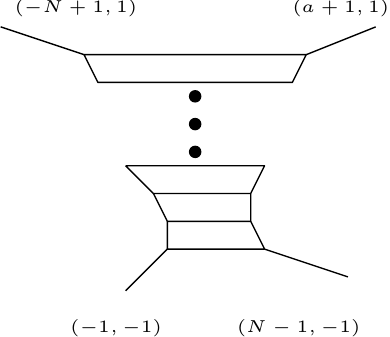}
\caption{The web dual to the Calabi-Yau threefold $X_{N,a}$. The K\"ahler parameters $t_{b}$ and $t_{f_{i}}$ are also indicated.}\label{CSweb}
\end{figure}

We can generalize the above to the case of $N>2$. The 5D $SU(N)$ gauge theory (with $N>2$) can be engineered using $N+1$ topologically distinct Calabi-Yau threefolds $X_{N,k}$ (with $k=0,1,\cdots,N$). In the field theory these Calabi-Yau threefolds are distinguished by the Chern-Simons term \cite{MS, Intriligator:1997pq}. These distinct Calabi-Yau threefolds have web diagrams \cite{Aharony:1997bh} which are also distinct from each other and are shown in \figref{CSweb}.

Notice that this web does not have parallel external legs except for $a=0$. Thus only in this, Chern-Simons term equal to zero, case can we compactify the external legs to obtain partially or fully compact web. If we compactify two of the external legs so that the web lives on a cylinder then the total space is given by
\bea
\widetilde{\mathbb{C}^{2}/\mathbb{Z}_{N}}\times_{f}T^2\,.
\eea

The K\"ahler parameter associated with the base elliptic curve will be denoted by $\tau$ and the partition function will have modular properties with respect to this parameter.

Thus the dual Calabi-Yau threefold is ellipticlly fibered and can be used to engineer six dimensional theories using F-theory compactification. The partition function of this six dimensional theory on $\mathbb{R}^{4}\times T^2$ is given by the elliptic genus of the $SU(N)$ instanton moduli space \cite{Haghighat:2013gba}.

\begin{figure}[h]
  \centering
  \includegraphics[width=2.5in]{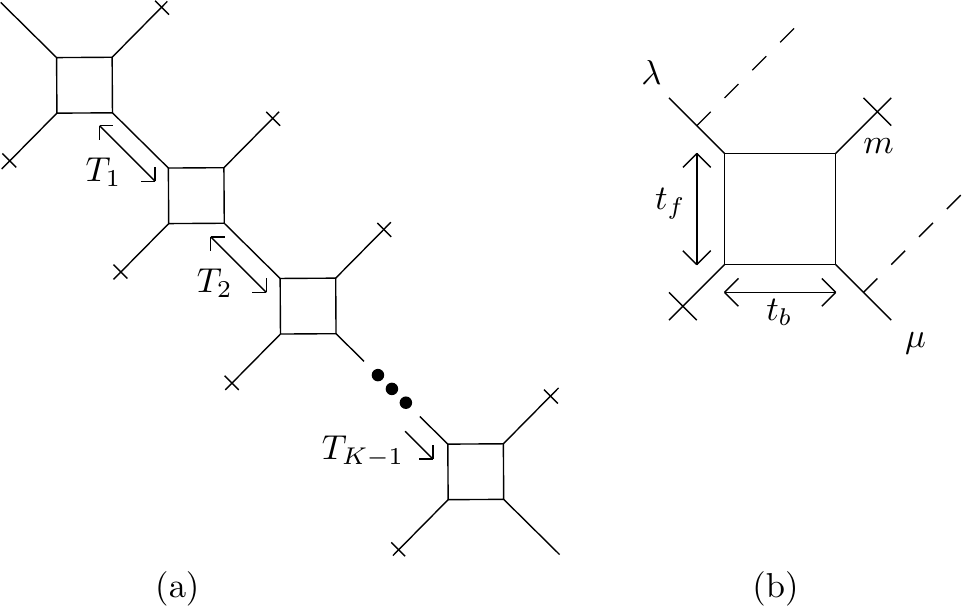}
\caption{(a) The compactification of web corresponding to product $SU(2)$ gauge theory. (b)The building block $G_{\lambda\mu}$ of the product $SU(2)$ web. The dashed lines indicate the Lagrangian branes with holonomy $\mbox{Tr}_{\lambda}V_{1}$ and $\mbox{Tr}_{\mu}V_{2}$.}\label{chain1}
\end{figure}

We can obtain a gauge theory with product gauge group $\prod_{i}SU(2)_{i}$ by gluing a chain of such local Calabi-Yau threefolds with dual web diagram as shown in \figref{chain1}(a). The building block of this chain is the local Calabi-Yau threefold with two Lagrangian branes as shown in \figref{chain1}(b).

\section{The Partition Function from Topological Vertex}
We can determine the partition function of the gauge theory engineered by the web in \figref{chain1}(a) by gluing together the open string amplitudes corresponding to the Lagrangian branes shown in \figref{chain1}(b). If we denote this open strings amplitude by $G^{(2)}_{\lambda\mu}$ then the partition function of the chain \figref{chain1} is given by,
\bea\nn
{\cal Z}_{2}=\sum_{\vec{\lambda}}(-Q_{1})^{|\lambda_{1}|}\cdots (-Q_{K-1})^{|\lambda_{K-1}|}\,G_{\emptyset\lambda_{1}}G_{\lambda_{1}^{t}\lambda_{2}}\cdots G_{\lambda_{K-1}^{t}\emptyset}\,,
\eea
where $K$ is the number of 4-cycles glued together and $Q_{i}=e^{-T_{i}}$. The open string amplitude $G^{(2)}_{\lambda\mu}$ is the building block of such partition functions and is a direct generalization of the building block or the ''domain wall" partition function studied in \cite{Haghighat:2013gba} and depends on the parameters $t_b, t_f$, $m$ and $\epsilon$, where $\epsilon$ is the Omega deformation parameter.



The partition function $G^{(2)}_{\lambda\mu}$ can be calculated using the topological vertex formalism in the unrefined limit. However, the refined case is more subtle since the refined topological vertex formalism \cite{IKV} requires the choice of a preferred direction for each of the vertices such that for all vertices in the web this direction should be the same (preferred directions for all vertices should be parallel). This may or may not be possible for a given web and therefore the refined vertex formalism only applies to a certain class of web diagrams. A well known example to which the refined vertex formslism can not be applied is the web corresponding to local $\mathbb{P}^2$ which was discussed in detail in \cite{Iqbal:2012mt}. We should note that this does not imply that refinement can not be done for the webs for which we can not find a set of parallel preferred directions for all the vertices, this is simply an issue with the formalism. For example, the refined partition function of the local $\mathbb{P}^2$ can be calculated using other methods including the refinement of the B-model \cite{Huang:2011qx}. Hence, for the refined case if the Lagrangian branes are on the preferred direction then there is no choice of the preferred direction for the upper right and lower left vertex (\figref{chain1}(b)). Hence, refined vertex formalism can not be used. In this case the new topological vertex discussed in \cite{Iqbal:2012mt} has to be used. Fortunately, for the product gauge group we are interested in the preferred direction can be the two horizontal lines which cover all four vertices. But we restrict ourselves to the unrefined case so that the building block partition function is given by,
\bea\nn
&&G^{(2)}_{\lambda\mu}(t_{b},t_{f},m,\epsilon)=\sum_{\nu_{1,2}\sigma\delta_{1,2}}(-Q_{b})^{|\nu_{1}|+|\nu_{2}|}(-Q_m)^{|\sigma|}\\\nn&\times& 
(-Q_{f})^{|\delta_{1}|+|\delta_{2}|}
C_{\delta_{1}\nu_{1}\lambda}(q)\,C_{\nu_{1}^{t}\delta_{2}\sigma}(q)\,C_{\delta_{2}^{t}\nu_{2}\mu}(q)\,C_{\nu_{2}^{t}\delta_{1}^{t}\sigma^{t}}(q)\,,
\eea 
where $Q_{b}=e^{-t_{b}}$, $Q_{f}=e^{-t_{f}}$, $Q_{m}=e^{-m}$. The topological vertex is given by ($q=e^{2\pi i \epsilon}$),
\bea\nonumber
C_{\lambda\,\mu\,\nu}(q)=q^{\frac{\kappa(\mu)}{2}}\,s_{\nu^t}(q^{-\rho})
\sum_{\eta}\,s_{\lambda^{t}/\eta}(q^{-\rho-\nu})\,s_{\mu/\eta}(q^{-\nu^{t}-\rho})
\eea
where $s_{\lambda/\eta}(\mathbf{x})$ is the skew-Schur function,
$q^{-\nu-\rho}=\{q^{-\nu_{1}+\tfrac{1}{2}},q^{-\nu_{2}+\tfrac{3}{2}},\cdots\}$ and $\kappa(\nu)=\sum_{(i,j)\in \nu}(j-i)$ defines the framing factor. Using the above definition of the vertex and the standard Schur function identities we get,
\bea\label{newequation}
\frac{G_{\lambda\mu}}{G_{\emptyset\emptyset}}&=&\sum_{\sigma}^{\infty}Q_{m}^{|\sigma|}Z_{\lambda\mu\sigma}(t_{b},t_{f},m,\epsilon)\\\nn
Z_{\lambda\sigma\mu}&=&\prod_{(i,j)\in \lambda}
\frac{\theta_{1}(t_{b}+t_{f}-2m;a_{ij})}{\theta_{1}(t_{b}+t_{f}-2m;b_{ij})}\\\nn\times
&&\prod_{(i,j)\in \sigma}\frac{\theta_{1}(t_{b}+t_{f}-2m;c_{ij})}{\theta_{1}(t_{b}+t_{f}-2m;d_{ij})}\frac{\theta_{1}(t_{b}+t_{f}-2m;e_{ij})}
{\theta_{1}(t_{b}+t_{f}-2m;f_{ij})}\\\nn\times
&&\prod_{(i,j)\in \mu}
\frac{\theta_{1}(t_{b}+t_{f}-2m;g_{ij})}{\theta_{1}(t_{b}+t_{f}-2m;h_{ij})}
\eea
where
\bea\label{parameters11}
a_{ij}&=&-t_{f}+m-\epsilon(\lambda_{i}-j+\sigma^{t}_{j}-i+1)\\\nn
b_{ij}&=&\lambda_{i}-j+\lambda^{t}_{j}-i+1\\\nn
c_{ij}&=&-t_{f}+m+\epsilon(\sigma_{i}-j+\lambda^{t}_{j}-i+1)\\\nn
d_{ij}&=&\sigma_{i}-j+\sigma^{t}_{j}-i+1\\\nn
e_{ij}&=&-t_{f}+m-\epsilon(\sigma_{i}+\mu^{t}_{j}-i+1)\\\nn
f_{ij}&=&\epsilon(\sigma^{t}_{j}-i+\sigma_{i}-j+1)\\\nn
g_{ij}&=&-t_{f}+m+\epsilon(\mu_{i}-j+\sigma^{t}_{j}-i+1)\\\nn
h_{ij}&=&\epsilon(\mu^{t}_{j}-i+\mu_{i}-j+1)
\eea
and ($Q_{\tau}=e^{2\pi i \tau}$)
\bea\nn
\theta_{1}(\tau;z)=-iQ_{\tau}^{1/8}y^{1/2}\prod_{k=1}^{\infty}(1-Q_{\tau}^k)(1-y\,Q_{\tau}^k)
(1-y^{-1}\,Q_{\tau}^k)
\eea
is the Jacobi theta function satisfying,
\bea\nn
\theta_{1}(\tau+1;z)=\theta_{1}(\tau;z)\,,\,\,\,\theta_{1}\Big(-\tfrac{1}{\tau};\tfrac{z}{\tau}\Big)=
-i(-i\tau)^{\frac{1}{2}}e^{\frac{i\pi z^2}{\tau}}\theta_{1}(\tau;z)\,.
\eea
It is clear from Eq.(\ref{newequation}) that the modular parameter in this case is $t_{b}+t_{f}-2m$. The partition function $Z_{\lambda\sigma\mu}$ is not modular invariant but transforms in the following way,
\bea\label{transformation11}
Z_{\lambda\sigma\mu}\Big(-\tfrac{1}{\tau},\tfrac{t_{f}}{\tau},\tfrac{m}{\tau},\tfrac{\epsilon}{\tau}\Big)=e^{\frac{i\pi r}{\tau}}\,Z_{\lambda\sigma\mu}(\tau,t_{f},m,\epsilon)
\eea
where $r$ is a quadratic function of the parameters given in Eq.(\ref{parameters11}). Note that we traded $t_{b}$ for $\tau$ and have taken $(\tau,t_{f},m)$ as the independent set of parameters. Recall that the web shown in \figref{su2webs}(a) gives rise to $SU(2)$ gauge theory with $t_{b}$ beig the gauge coupling and $t_{f}$ the Coulomb branch parameter (vev of the scalar breaking the $SU(2)$). Additional compactification of the two external legs gives a web which is related to the $N=2^{*}$ gauge theory with coupling constant $\tau$ as discussed in detail in \cite{Haghighat:2013gba}. Thus in choicing the independent set of parameters we simply choice the parameters preferred in the gauge theory: the coupling constant, the Coulomb branch parameter and the mass of the adjoint.

We can make $Z_{\lambda\sigma\mu}$ invariant under the modular transformation at the expense of introducing the holomorphic anomaly \cite{Bershadsky:1993ta,BCOV,Witten:1993ed, Alim:2010cf}. Recall that the Jacobi theta function $\theta_{1}(\tau,z)$ has the following representation in terms of the Eisenstein series $E_{2k}(\tau)$,
\bea\nn
\theta_{1}(\tau;z)=\eta^{3}(\tau)(2\pi iz)\mbox{exp}\Big(\sum_{k\geq 1}\frac{B_{2k}}{(2k)(2k)!}E_{2k}(\tau)(2\pi iz)^{2k}\Big)\,.
\eea
The exponential factor in Eq.(\ref{transformation11}) is due to the presence of $E_{2}(\tau)$ in the above expression of $\theta_{1}(\tau;z)$ since $E_{2}(\tau)$ is transforms in the following way under the modular transformation,
\bea
E_{2}\Big(-\frac{1}{\tau}\Big)=\tau^{2}E_{2}(\tau)-i\pi \tau\,.
\eea
We can replace $E_{2}(\tau)$ with $E_{2}(\tau,\overline{\tau})=E_{2}(\tau)-\frac{3}{Im(\tau)}$ which transforms as a weight two modular form but is not holomorphic. This replacement leads to a holomorphic anomaly which gives:
\bea
Z_{\lambda\sigma\mu}\Big(-\tfrac{1}{\tau},\tfrac{t_{f}}{\tau},\tfrac{m}{\tau},\tfrac{\epsilon}{\tau}\Big)&Z&_{\lambda\sigma\mu}(\tau,t_{f},m,\epsilon)\,,\\\nonumber
\,\,\,\frac{\partial Z_{\lambda\sigma\mu}(\tau,t_{f},m,\epsilon)}{\partial E_{2}(\tau,\overline{\tau})}&=&\frac{1}{24}\,r\,Z_{\lambda\sigma\mu}(\tau,t_{f},m,\epsilon)\,.
\eea
Similarly we can generalize this result to the case of a web made of pieces which are $\mathbb{Z}_{N}$ orbifold of the $N=1$ case. This is shown in \figref{chain3} below.

\begin{figure}[h]
  \centering
  \includegraphics[width=3.5in]{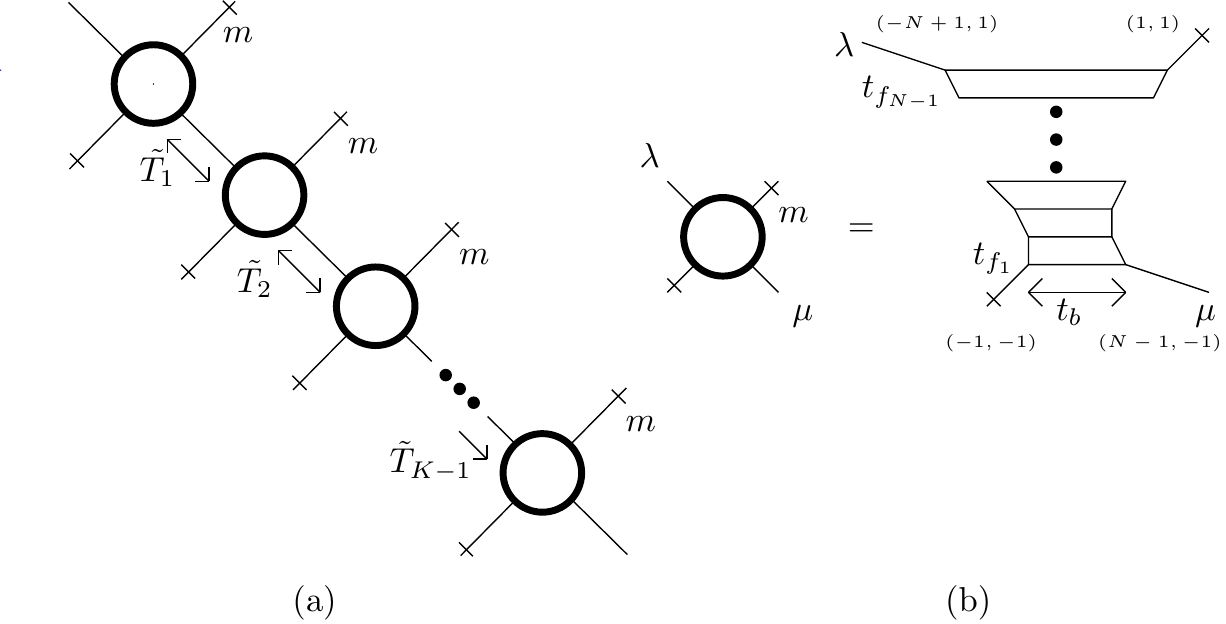}
\caption{The compactfication of the $X_{N,0}$ web.}\label{chain3}
\end{figure}

The partition function of this chain of $SU(N)$ webs is given by
\bea
{\cal Z}_{N}=\sum_{\vec{\lambda}}(-\tilde{Q}_{1})^{\lambda_{1}}\cdots \tilde{Q}_{K-1}^{|\lambda_{K-1}|}G_{\emptyset\lambda_{1}}\cdots G_{\lambda^{t}_{K-1}\emptyset}
\eea
where $\tilde{Q}_{s}=e^{-\tilde{T}_{s}}$, $s=1,\cdots, K-1$ and $G_{\lambda\mu}$ can again be expressed in terms of Jacobi theta function with modular parameter $t_{b}-\sum_{i=1}^{N-1}t_{f_{i}}-N\,m$ where as discussed before $t_{f_{i}}$ are the K\"ahler parameters of the fiber $\mathbb{P}^{1}$'s.

\section{Conclusions}
In this paper we have studied some new compactified webs which lead to gauge theories with product gauge group of type $\prod_{a}SU(N)_{a}$. We worked out the partition function of this web for the case $N=2$ explicitly and showed that it can be written in terms of building blocks which are generalization of the M-string domain wall partition functions (which correspond to $N=1$). This building block for the case $N=2$ has interesting modular properties and satisfy a holomorphic anomaly equation similar to the case of $N=1$ partition function have 

It will be interesting to see if this partition function can be determined using index calculus on some instanton moduli spaces \cite{Nekrasov,HIV, nakajima}. This kind of description can allow one to see if the partition function can be expressed as the $(2,2)$ or $(0,2)$ elliptic genus of some target space.

\noindent\section*{ Acknowledgments}
KS would like to thank Amer Iqbal for many useful discussions.

\bibliography{physics}

\end{document}